\documentclass{aa}

\usepackage{graphicx}
\usepackage{rotating}
\usepackage{txfonts}

\def\dsct{$\delta$~Scuti }
\def\Msun{$M_{\odot}$ }

\def\cd{~d$^{\rm -1}$}

\begin{document}

\title{Regular frequency patterns in the classical $\delta$ Scuti star HD~144277 observed by the MOST\thanks{Based on data from the {\it MOST} satellite, a Canadian Space Agency mission, jointly operated by Microsatellite Systems Canada Inc. (MSCI), formerly part of Dynacon, Inc., the University of Toronto Institute for Aerospace Studies, and the University of British Columbia with the assistance of the University of Vienna.} satellite}

\author{K. Zwintz\inst{1} \and
P. Lenz\inst{2} \and
M. Breger\inst{1,3} \and
A. A. Pamyatnykh\inst{1,2,4} \and
T. Zdravkov\inst{2} \and
R. Kuschnig\inst{1} \and
J. M. Matthews\inst{5} \and
D. B. Guenther\inst{6} \and
A. F. J. Moffat\inst{7} \and
J. F. Rowe\inst{8} \and
S. M. Rucinski\inst{9} \and
D. Sasselov\inst{10} \and
W. W. Weiss\inst{1}
}

\offprints{K. Zwintz, \\ \email{konstanze.zwintz@univie.ac.at}}

\institute{
   Institut f\"ur Astronomie, Universit\"at Wien,
    T\"urkenschanzstrasse 17, A-1180 Vienna, Austria \\
    \email firstname.lastname@univie.ac.at  \and
   Copernicus Astronomical Centre, Bartycka 18, 00-716 Warsaw, Poland \and
   Dept. of Astronomy, University of Texas at Austin, Austin, TX 78712, USA \and
   Institute of Astronomy, Russian Academy of Sciences, Pyatnitskaya Str 48, 109017 Moscow, Russia \and
   Department of Physics and Astronomy, University of British Columbia,
    6224 Agricultural Road, Vancouver, BC V6T 1Z1, Canada \and
    Department of Astronomy and Physics, St. Mary's University, Halifax,
    NS B3H 3C3, Canada \and
    D\'epartment de physique, Universit\'e de Montr\'eal, C.P.6128, Succ. Centre-Ville, Montr\'eal, QC H3C 3J7, Canada \and
    NASA-Ames Research Park, MS-244-30, Moffett Field, CA 94035, USA \and
    David Dunlap Observatory, University of Toronto, P.O. Box 360,
    Richmond Hill, ON L4C 4Y6, Canada \and
    Harvard-Smithsonian Center for Astrophysics, 60 Garden Street,
    Cambridge, MA 02138, USA
}

\date{Received / Accepted }

\abstract
{We present high-precision time-series photometry of the classical \dsct star HD 144277 obtained with the MOST (Microvariability and Oscillations of STars) satellite in two consecutive years. The observed regular frequency patterns are investigated asteroseismologically.}
{HD~144277 is a hot A-type star that is located on the blue border of the classical instability strip. While we mostly observe low radial order modes in classical \dsct stars, HD~144277 presents a different case. Its high observed frequencies, i.e., between 59.9\cd (693.9\,$\mu$Hz) and 71.1\cd (822.8\,$\mu$Hz), suggest higher radial orders. We examine the progression of the regular frequency spacings from the low radial order to the asymptotic frequency region. }
{Frequency analysis was performed using Period04 and SigSpec. The results from the MOST observing runs in 2009 and 2010 were compared to each other. The resulting frequencies were submitted to asteroseismic analysis. }
{HD 144277 was discovered to be a \dsct star using the time-series photometry observed by the MOST satellite. Twelve independent pulsation frequencies lying in four distinct groups were identified. Two additional frequencies were found to be combination frequencies. The typical spacing of 3.6\cd corresponds to the spacing between subsequent radial and dipole modes, therefore the spacing between radial modes is twice this value, 7.2\cd. Based on the assumption of slow rotation, we find evidence that the two radial modes are the sixth and seventh overtones, and the frequency with the highest amplitude can be identified as a dipole mode. }
{The models required to fit the observed instability range need slightly less metallicity and a moderate enhancement of the helium abundance compared to the standard chemical composition. Our asteroseismic models suggest that HD 144277 is a \dsct star close to the ZAMS with a mass of 1.66\,\Msun. }


\maketitle
\titlerunning{HD~144277}
\authorrunning{K. Zwintz et al.}

\section{Introduction}

The extensive photometric campaigns of \dsct stars are producing important information on the values and amplitudes of the excited pulsation frequencies. However, mode identifications are required for unique theoretical modeling. These mode identifications can be obtained from spectroscopic line-profile variations, as well as from photometric color information from the light curves. However, these techniques usually can only be applied to a few of the detected frequencies. Consequently, the recognition of regular frequency patterns in the power spectra becomes an important additional tool. 

Statistical analyses of the frequency spacings of $\delta$ Scuti stars show that in many stars the photometrically observed frequencies cluster around the frequencies of the radial modes over many radial orders (Breger, Lenz \& Pamyatnykh 2009).
The observed regularities can be partly explained by modes trapped in the stellar envelope. In particular, the low-order $\ell$ = 1 modes have frequencies close to those of the radial modes. We call this the low-order spacing, which means that the nearly regular frequency patterns correspond to successive radial orders, which can then be used to infer the stellar $\log g$ values (Breger, Lenz \& Pamyatnykh 2008).

This contrasts with the asymptotic case, where the $\ell$ = 1 frequencies are found nearly halfway between successive radial orders of the radial and $\ell$ = 2 modes. In the asymptotic case, successive frequencies (and frequency clusters) correspond to alternating $\ell$ values. The observed regularities differ from the low-order situation by a factor of two.

We note here that regular frequency spacings can also be caused by combination modes. In extreme cases the combination modes cover the whole frequency range from the low-order frequencies up to the asymptotic frequency range. An impressive example of such a star is KIC 9700322 (Breger et al. 2011) measured by the $Kepler$ satellite. However, close inspection of the frequency patterns can uniquely identify the combination modes. We do that in this paper.

\begin{figure}[htb]
\centering
\includegraphics[width=0.5\textwidth]{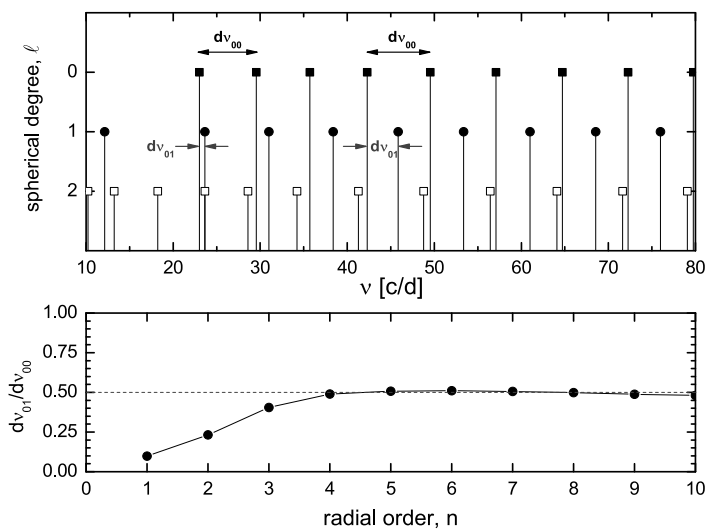}
\caption{Theoretically computed frequencies of $\ell$ = 0, 1, and 2 modes as a function of radial order. At low order, the $\ell$ = 1 modes have frequencies near those of radial modes. As the radial order increases, the $\ell$ = 1 modes move towards positions approximately halfway between successive orders of the $\ell$ = 0, 2 modes (asymptotic relation). This explains the frequency regularities seen in many $\delta$ Scuti stars, which seemingly are at odds with the asymptotic relation by a factor up to 2.}
\label{intro}
\end{figure}

A consequence of the different behavior between the low-order and asymptotic cases is that the photometrically detectable frequency spacings differ by a factor two. The `wandering' of the $\ell$ = 1 modes is illustrated in Figure \ref{intro}, which was computed from a nonrotating ZAMS model of 1.8\,\Msun The corresponding diagrams using observations are not yet available because for $\delta$ Scuti stars, $\ell$ = 1 mode identifications are available only for low-order modes.

To observationally examine the progression (doubling) of the regular frequency spacings from the low-radial order to the asymptotic
frequency region, it is necessary to look at stars pulsating in high orders close to the asymptotic case. Such stars can be found near the hot border of
the $\delta$ Scuti instability strip (Pamyatnykh \cite{pam00}), which was already suspected thirty years ago.
HD~144277 is such a candidate.

Very little is known about HD~144277 ($V$ = 7.88 mag) in the literature. It has a spectral type of A1V with a respective effective temperature of 9230~K (both taken from the Tycho-2 Spectral Type Catalog; Wright et al. \cite{wri03}) and a parallax of 7.00$\pm$5.59 mas (Kharchenko et al. \cite{kha09}).

To confirm the location near the blue border of the classical instability strip, Str\"omgren colors for HD~144277 were obtained with the South African Astronomical Observatory (SAAO) 50cm telescope on 2010 February 1. We derived the following values from these measurements: \mbox{$b-y~=~0.036$ mag}, \mbox{$m_1$~=~0.186 mag} and $c_1$~=~0.985 mag. This confirms that HD~144277 is a hot A star.

\section{MOST observations}
The MOST space telescope (Walker et al. \cite{wal03}) orbits the Earth in its polar Sun-synchronous circular orbit at an altitude of 820\,km with a period of 101.413\,minutes. Since its launch on 30 June 2003, the satellite has observed numerous objects with its 15-cm Rumak-Maksutov telescope which feeds a CCD photometer through a single, custom broadband filter (wavelength range from 350 to 750\,nm).

MOST can provide three types of photometric data simultaneously for different targets in its field of view. Data obtained in Fabry Imaging Mode are generated by a projection of the entrance pupil of the telescope -- illuminated by a bright ($V < 6$ mag) target star -- onto the Science CCD by a Fabry microlens (see Reegen et al. \cite{ree06} for details). Direct Imaging is used for stars in the open area of the CCD that is not covered by the Fabry microlens-array field-stop mask. It resembles conventional CCD photometry where photometry is obtained from defocussed images of stars. Although originally not intended for scientific purposes, the MOST Guide Stars used for the Attitude Control System (ACS) provide highly accurate photometry, which has been frequently used in the past (e.g., Zwintz et al. \cite{zwi09}). 

HD~144277 was observed in Guide Star photometry mode. It lies slightly outside the MOST continuous viewing zone, which means it can be observed only for a part of each 101-min orbit.

MOST observed HD~144277 from 2009 April 18 to 30 for 12.03 days with a duty cycle of 39\% and from 2010 May 3 to 25 for 21.48 days with a duty cycle of 41\%. (The duty cycle represents the fraction of each 101-min orbit covered by the data; there are no long gaps in the time-series.) In both years onboard exposures were 1\,s long (to satisfy the cadence of guide star ACS operations). In 2009, 61 consecutive exposures were ``stacked'' onboard to produce integrations 61\,s long, while in 2010 the integration times were 30\,s as 30 consecutive images were used.

\begin{figure}[htb]
\centering
\includegraphics[width=0.45\textwidth]{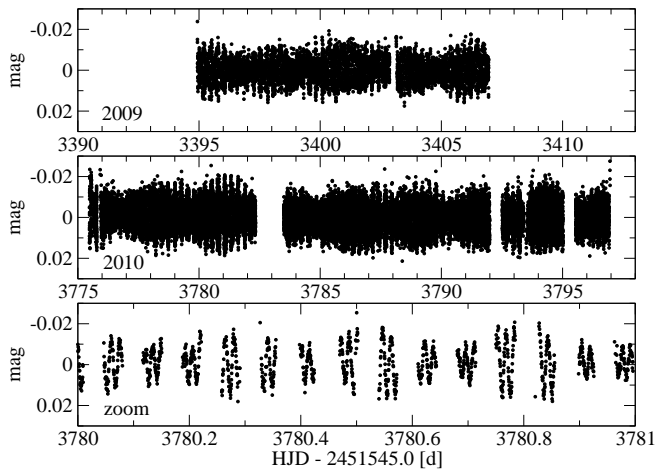}
\caption{MOST time-series photometry of HD~144277: 2009 observations (top panel) and 2010 observations (middle panel) to the same scale. A zoom into the 2010 light curve shows the pulsational variability (bottom panel).}
\label{lcs}
\end{figure}

The respective light curves from 2009 and 2010, together with a zoom into the 2010 data, are shown in Figure \ref{lcs}. Using the data from the MOST space telescope the \dsct nature of HD~144277 was discovered.

\section{Data reduction and frequency analysis}
The data for HD~144277 were obtained in MOST Guide Star photometry mode. For this type of data, a special reduction method was developed (see Hareter et al. \cite{har08} for details), which uses a similar approach as for targets observed in Fabry Imaging Mode (Reegen et al. \cite{ree06}), i.e., resolving linear correlations between the intensity of target and background pixels.  Since MOST Guide Star photometry does not provide information on the intensity of the background, the correlations between constant and variable stars are used instead. In a first step, the MOST Guide Stars are classified into variable and constant objects by a quick-look analysis. The data of all selected intrinsically constant objects are then combined to a comparison light curve.

Stray light effects are corrected by subtraction of linear correlations between target and comparison time-series. It is self-evident that comparison time-series are assumed and required to contain no variable stellar signal. If one of the stars used for the comparison light curve turns out to be variable (even at low amplitude levels) in a later stage of the analysis, the complete reduction has to be repeated omitting the variable star. 
After the Guide Star photometry reduction, the 2009 light curve of HD~144277 contains 6366 data points, the 2010 light curve has 23061 data points corresponding to Nyquist frequencies of 705\cd\, and 1411\cd, respectively. 

\begin{figure}[htb]
\centering
\includegraphics[width=0.45\textwidth]{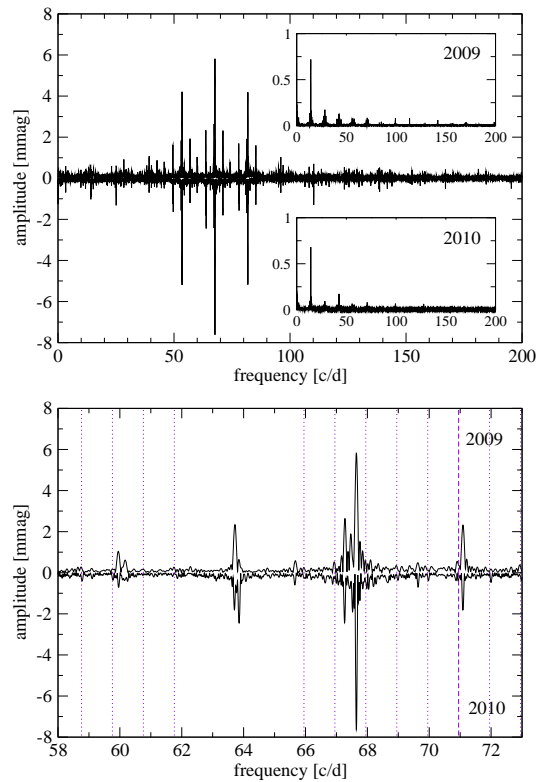}
\caption{Amplitude spectra of HD~144277 from 2009 (pointed upwards) and 2010 (pointed downwards); {\it top panel}: complete spectra from 0 to 200\cd where the respective spectral windows are given as insets; {\it bottom panel}: zoom into the region from 58 to 73\cd where the associated multiple of the MOST orbital frequency is shown as dashed line and the 1\cd sidelobes as dotted lines.}
\label{amps}
\end{figure}

Frequency analysis was conducted using the Period04 software (Lenz \& Breger \cite{len05}) where Fourier and least-squares algorithms are combined. SigSpec (Reegen \cite{ree07}) was used to verify the results. A signal was considered to be significant if its amplitude signal-to-noise (S/N) value was to be larger than 4.0 (Breger et al. \cite{bre93}, Kuschnig et al. \cite{kus97}), and the SigSpec significance was higher than the respective threshold (Reegen \cite{ree07}), which is computed as

\begin{equation}
siglimit = sig + {\rm log}(K/2)
\end{equation}
where $siglimit$ is the threshold for the significances, $sig$ is the significance of a given peak, and $K$ the number of data points. For the 2009 MOST data set the threshold $siglimit$ accounts to 8.96 and for the 2010 data set it is 9.52. The amplitude spectra for both data sets (2009 pointed upwards and 2010 pointed downwards) are shown in Figure \ref{amps} including a zoom in the region where the pulsation frequencies were discovered.

\begin{figure}[htb]
\centering
\includegraphics[width=0.45\textwidth]{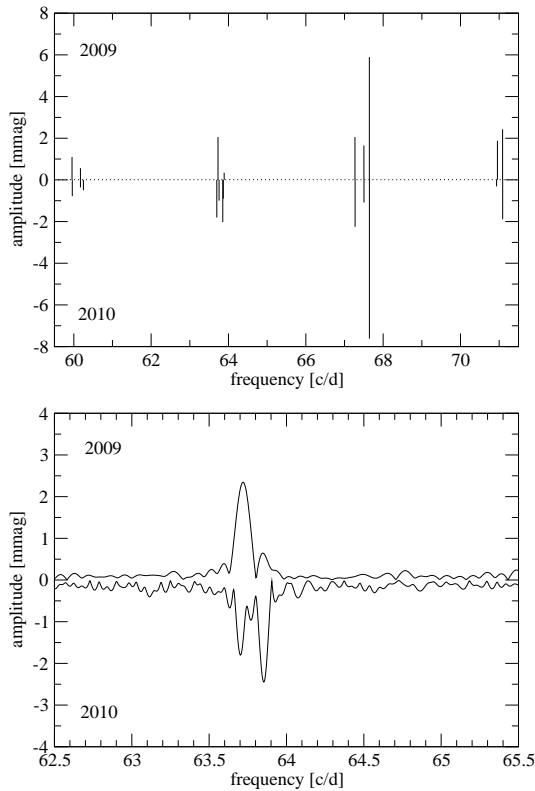}
\caption{Pulsation frequencies of HD~144277 where data from 2009 are pointed upwards and data from 2010 are pointed downwards. {\it Top panel}: Pulsation frequency spectra. (Errors in frequency are not visible because they are as small as the thickness of the lines.) {\it Bottom panel}: Zoom into the amplitude spectrum from 62.5 to 65.5\cd ilustrating the higher frequency resolution of the 2010 data and the actual frequency quadruplet around 63.8\cd. }
\label{fspec}
\end{figure}

The analysis with Period04 yielded 28 formally significant frequencies with S/N values higher than 4.0 for the data set from 2009 and 33 for the 2010 light curve both calculated from 0 to 200\cd. Applying the significance thresholds given above, SigSpec finds 43 formally significant frequencies for the 2009 data and 51 peaks for the 2010 data.
All formally significant frequencies were then checked against instrumental frequencies related to the orbit of the satellite, its harmonics and 1\cd sidelobes within the frequency resolution (computed according to Kallinger et al. \cite{kal08}). All peaks related to instrumental frequencies were discarded in the analysis. In total, 11 peaks were attributed to pulsation in the data set from 2009. All of them were confirmed by the 2010 data set, and an additional three were discovered due to the longer time-series, hence lower noise level. The respective pulsation frequency spectra are plotted in the top panel of Figure \ref{fspec}. Although the frequency errors are also indicated, they are mostly smaller than the respective thickness of the lines.

Table \ref{freqs} lists the pulsation frequencies identified in the 2009 and 2010 data with their respective last digit errors given in parenthesis computed according to Kallinger et al. (\cite{kal08}). The numbering is given according to increasing frequency values. Thanks to the longer observations, the accuracy of the 2010 data is higher and allows detecting additional frequencies at lower amplitudes. Frequency F13 at 131.03\cd  and F14 at 138.36\cd are combination frequencies within the given uncertainties.
A closer look at the frequencies around 63.7\cd shows that in the 2009 data only two frequencies could be resolved in this group (B), i.e., F4 and F7. But with the longer time-series from 2010 it is shown that in fact this is a frequency quadruplet (see Figure \ref{fspec}).

\begin{figure}[htb]
\centering
\includegraphics[width=0.45\textwidth]{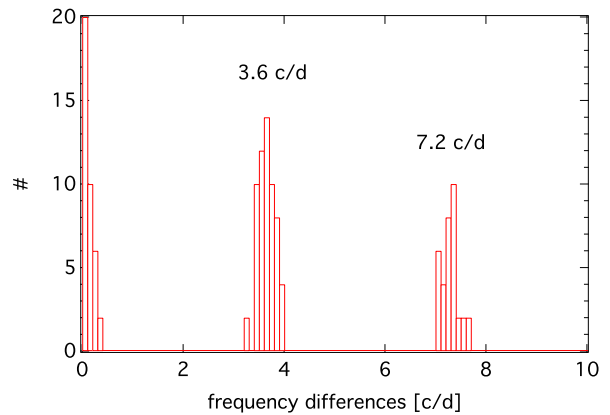}
\caption{Histogram of the frequency differences of the observed modes computed using the 2010 data set of HD~144277.}
\label{histo}
\end{figure}

From Figures \ref{amps} and \ref{fspec}, it is evident that there are four distinctive groups of frequencies with nearly equidistant spacing (marked A to D in Table \ref{freqs}). A histogram of all observed frequency differences (see Figure \ref{histo}) shows that the spacing between the frequency groups is 3.6\cd (or 41.67\,$\mu$Hz). This is not a typical phenomenon for classical \dsct stars. But Breger, Lenz \& Pamyatnykh (\cite{breger09}) report such regularities in the \dsct stars 44 Tau, FG Vir, and BL Cam. The pulsation frequencies of these three objects lie at much lower values than those of HD~144277. Additionally, the stars are cooler than HD~144277 and seem to be more evolved. Inserting HD~144277's observed properties (i.e. a separation of 3.6\cd and an assumed radial mode of 60-70\cd) into the theoretical $s-f$ diagram of Breger, Lenz \& Pamyatnykh (\cite{breger09}, their Figure 8) already suggests high overtone modes that are confirmed by our asteroseismic models.

\begin{table*}[htb]
\caption{Pulsation frequencies of HD144277 identified in both data sets (2009 and 2010), amplitudes, significances, amplitude signal-to-noise values (S/N), and the respective frequency errors given in parentheses.}
\label{freqs}
\begin{center}
\begin{footnotesize}
\begin{tabular}{lllrrrllrrrc}
\hline
\multicolumn{1}{c}{\#} & \multicolumn{5}{c}{2009 data}  & \multicolumn{5}{c}{2010 data} & \multicolumn{1}{c}{ }  \\
\hline
 & \multicolumn{2}{c}{freq} & \multicolumn{1}{c}{amp} & \multicolumn{1}{c}{sig} & \multicolumn{1}{c}{S/N}
 & \multicolumn{2}{c}{freq} & \multicolumn{1}{c}{amp} & \multicolumn{1}{c}{sig}  & \multicolumn{1}{c}{S/N} & \multicolumn{1}{c}{Group} \\
 & \multicolumn{1}{c}{[$d^{-1}$]} & \multicolumn{1}{c}{[$\mu$Hz]} & \multicolumn{1}{c}{[mmag]} &  &  & \multicolumn{1}{c}{[$d^{-1}$]} & \multicolumn{1}{c}{[$\mu$Hz]} & \multicolumn{1}{c}{[mmag]} & & &   \\
\hline
F1 &	59.954(7) &	693.91 (8) & 	1.097 & 	138.24  & 	11.2 &	59.962(4) & 	694.00(4)  & 	0.780 & 	162.31 &  10.5 & A \\
F2 &	60.170(10)  &	696.40(20)  & 	0.564 & 	37.69 & 6.7 &		60.172(7) & 	696.43(8)  & 	0.364 & 	42.76  & 7.8 & A \\
F3 &			&			&		&		&	&  		60.248(5) & 	697.32(6)  & 	0.496 & 	76.08  & 8.1 & A \\
F4 &	63.728(5) &	737.60(6) & 	2.054 & 	278.30 & 26.4 &	63.699(2) & 	737.25(2)  & 	1.807 & 	649.80  & 26.1 & B  \\
F5 &			&			&		&		&	&		63.756(3) & 	737.92(4)  & 	0.994 & 	203.10  & 23.0 & B\\
F6 &			&			&		&		&	&		63.852(2) & 	739.03(2)  & 	2.020 & 	793.51  & 20.5 & B\\
F7 &	63.890(20)  &	739.50(20)  & 	0.343 & 	19.73 & 5.8 &		63.865(7) & 	739.18(8)  & 	0.902 & 	50.12  & 8.2 & B \\
F8 &	67.271(5)  &	778.60(5) & 	2.055 & 	327.10 & 	12.1 &	67.274(2) & 	778.63(2)  & 	2.250 & 	806.11  & 27.8 & C \\
F9 &	67.503(6)  &	781.28(7) & 	1.650 & 	190.00 & 23.9 &	67.504(3) & 	781.30(3)  & 	1.087 & 	279.10  & 20.5 & C \\
F10 &	67.643(3)  &	782.90(4) & 	5.893 & 	709.71 & 31.0 &		67.644(1) & 	782.91(1)  & 	7.616 & 	2992.07  & 52.6 & C \\
F11 &	70.960(10) &	821.20(20) & 	1.875 & 	29.63 & 6.8 &		70.930(10)   & 	821.00(20)   & 	0.310 & 	9.82  & 6.6 & D \\
F12 &	71.089(5) &	822.79(6) & 	2.432 & 	243.55 & 22.1 &		71.091(2) & 	822.81(2) & 	1.890 & 	617.41 & 23.2 &  D \\	
F13 &	131.040(30)  &	1516.60(30) & 	0.204 & 	9.46	 & 3.8 &		131.030(20) & 	1516.60(20) & 	0.215 & 	8.93  & 6.2 & F4 + F8 \\	
F14 &	138.360(20) &	1601.30(20) & 	0.470 & 	29.16 & 11.7 &		138.362(5) & 	1601.41(6) & 	0.516 & 	80.90  & 16.2 & F8 + F12 \\
\hline
\end{tabular}
\end{footnotesize}
\end{center}
\end{table*}

\section{Asteroseismic modeling of HD~144277}

To model the observed frequency spectrum (Figure \ref{fspec} top panel), we used the Warsaw-New Jersey evolutionary code and Dziembowski's pulsation code in the same versions as in Lenz et al. (\cite{lep2010}). Our standard models are obtained with OPAL opacities (Iglesias et al. \cite{iglesias1996}) supplemented with low-temperature opacities provided by Ferguson et al. (\cite{ferg05}). Moreover, we used the Asplund et al. (\cite{asplund09}) solar element mixture (A09 hereafter) with the corresponding solar composition: X=0.74, Z=0.0134. 
A convective core overshooting parameter $d_{\rm{ov}}$=0.1 has been adopted for all models. 
The extent of convective overshooting from the core, however, has no influence on the purely acoustic spectrum detected for HD~144277.
Rotation is treated following the assumption of conservation of global angular momentum and a uniform rotation law.

\subsection{Interpretation of the observed frequency spacing}

The observed frequencies (Table \ref{freqs}) are situated between 59 and 71\cd, with a spacing of 3.6\cd between the different frequency groups. Since the radial fundamental mode cannot exceed 25\cd, even for \dsct stars close to the ZAMS, we certainly observe modes of high radial order. 

The predicted spacing between radial modes for a main sequence star close to the ZAMS and with a typical mass of 1.8\,M$_\odot$ is depicted in Figure \ref{intro}. The figure also shows the spacing between the radial modes and the adjacent axisymmetric $\ell=1$ mode in this purely acoustic spectrum. As can be seen, the dipole modes move from a location close to the radial modes to a location midway between radial modes with increasing radial order.
In most \dsct stars, we predominantly observe modes of low radial orders, where the observed spacing may correspond to the spacing of radial modes (see, e.g., Breger et al. \cite{breger09}). HD~144277 presents a different case. Drawing our attention to the higher radial orders, we have to interpret the observed spacing of 3.6\cd as the spacing between alternating radial and dipole modes, meaning that the radial spacing corresponds to twice this value, 7.2\cd, representing the asymptotic instead of the low-order case (see Introduction).

In asteroseismic analyses the stellar rotation rate is a key variable. Since the projected rotational velocity of HD~144277 is not known, the most promising method of determining the rotation rate originates in the (potential) pattern of rotationally split modes. The observed frequency spectrum only leaves a few possibilities:

(i) rapid rotation and interpretation of the four groups of frequencies (see Table~\ref{freqs} and Figure~\ref{fspec}) as components of a rotationally split mode. It has been shown by 2D calculations (Espinosa et al. \cite{espinosa2004}, Burke et al. \cite{burke2011}) that rapid rotation forces adjacent components (m-values) of multiplets to form close frequency pairs. For example, an $\ell = 2$ quintuplet may appear as two close frequency pairs and one additional component, i.e., three groups. However, in our frequency groups we observe more than two components and, moreover, a very symmetric spacing between the groups, which is difficult to explain in the framework of this hypothesis. Formally, for a star close to the ZAMS a rotation frequency of 3.6\cd\, is still below the critical breakup frequency, $\nu_{\rm breakup} = 1/(2\pi) \sqrt{GM/R^3}$. For example for a 1.8\,M$_\odot$ model at the ZAMS the critical breakup frequency is approximately 5.2\cd,

(ii) slow rotation and interpretation of the substructure within the frequency groups as rotational splitting. This would lead to a rotation rate of about 15.0 km s$^{\rm -1}$,

(iii) other regular patterns in the framework of rapid rotation.
Recent investigations devoted to examining of acoustic mode frequencies in fast rotators by a ray-based asymptotic theory predict regular frequency patterns (e.g., Lignieres et al. \cite{lig09,lig10} and Reese et al. \cite{rees09}). These predicted regularities may serve as an additional possibility for interpreting the observed frequency spectrum of HD~144277. Moreover, it has been shown by simulations based on synthetic frequency spectra (Lignieres et al. \cite{lig10}) that it is more likely that we find regular patterns in pole-on configurations, which in turn would imply that the measured $v \sin i$ would not necessarily exhibit high values.


In this paper we restrict our investigation to the case of slow rotation, for which our pulsation code is valid.

\subsection{Determination of the location of the radial modes}

Given a presumed radial frequency spacing of 7.2\cd, there are two possibilities where the radial modes can be located, in the frequency group at 60 and 67\cd (hereafter: Hypothesis~1) or in the groups at 64 and 71\cd (hereafter: Hypothesis~2). Since no mode identification is available, we rely completely on asteroseismic models to choose between these two possibilities.

\begin{figure}[h]
  \centering
  \includegraphics[width=9cm,bb=10 15 300 215]{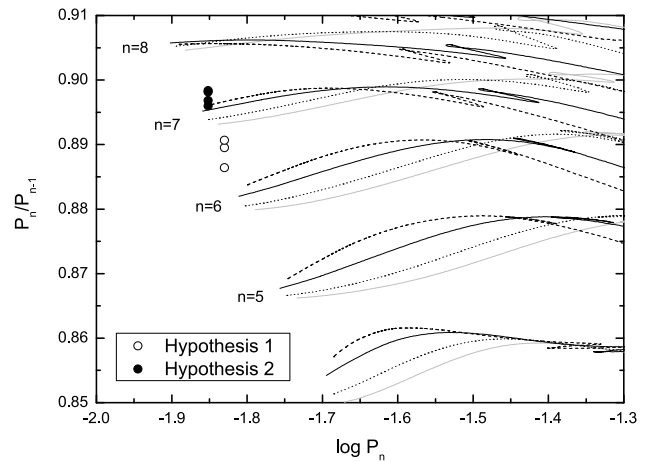}
  \caption{Period ratios of consecutive radial modes vs. the shorter periods for four models differing in mass: 1.5 (dashed), 1.8 (solid), 2.2 (dotted), and 2.5 M$_{\odot}$ (gray). The radial orders, $n$, of the shorter periods are labeled. The observational errors are smaller than the symbol size.}
  \label{fig:petersenmass}
\end{figure}

We performed an analysis based on a modified form of the Petersen diagram (Petersen \& J{\o}rgensen \cite{pet1972}); i.e., we plotted the predicted radial period ratios against the shorter period and compared the resulting tracks with the observations. The observational position in this diagram is different for the two hypotheses mentioned above. In the case of Hypothesis~1, we presume that F10 (i.e., the frequency with the highest amplitude) is a radial mode and compute its period ratios with the modes in group A. In the case of Hypothesis~2, the same is done for F12 (the dominant mode in group D) and the frequencies in group B.
The theoretical period ratios for four models of different masses (1.5, 1.8, 2.2, and 2.5\,M$_\odot$) are given in Figure~\ref{fig:petersenmass}. The empty circles denote the observational position following Hypothesis~1 and the filled circles correspond to Hypothesis~2. The results in Figure~\ref{fig:petersenmass} confirm the very young evolutionary stage of HD 144277 near the ZAMS. Within the given mass range the points derived with Hypothesis~2 are in better agreement with the models and indicate that we observe the sixth and seventh radial overtones. Hypothesis~1 leads to a period ratio that is in-between the predicted values, which means that the period ratio is not fitted simultaneously with the observed frequencies.

\begin{figure}[h]
  \centering
  \includegraphics[width=9cm,bb=10 15 300 215]{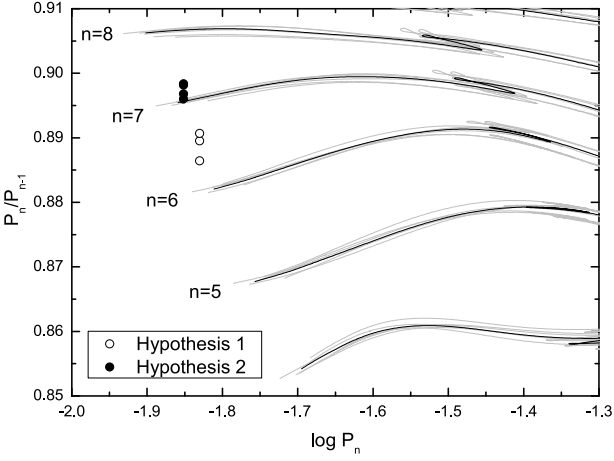}
  \caption{ Same as Fig.~\ref{fig:petersenmass}, but showing the effect of changing X, Z, rotation rate, and opacity data (depicted in gray) in comparison to a standard 1.8~M$_{\odot}$ model depicted with solid black lines. }
  \label{fig:petersen}
\end{figure}

Naturally the significance of this result needs to be checked, so, we examined the effect of changing various model parameters to study their influence on period ratios at high radial orders.
 In particular we varied the mass fraction of metal from the solar value Z\,=\,0.0134 to 0.010 and 0.020 respectively, changed the hydrogen mass fraction, X, from the solar value 0.74 to 0.70 and 0.78, varied the equatorial rotation velocity from 0 to 100\,kms$^{-1}$ (based on a second-order perturbation approach and neglecting near-degeneracy effects), and also examined the effect of OPAL vs. OP opacities (Seaton \cite{seaton05}). 

 The corresponding models are shown in Figure~\ref{fig:petersen} in comparison with a standard 1.8~M$_{\odot}$ model. At high radial orders, the effect of these parameter changes on the period ratio is very small close to the ZAMS, and Hypothesis 2 is still clearly preferred. In fact, the biggest effect on the high order radial period ratios besides mass change would come from near-degeneracy effects at fast rotation. However, these effects would be seen as a disturbance of the symmetric spacing in the frequency spectrum. This is obviously not observed. Our investigation focuses on the explanation of the observed frequency spectrum in the case of slow rotation, for which near-degeneracy effects would be very small. 
We attempted to fit the observed frequencies and their spacing according to Hypothesis 1 taking into account all combined effects, but did not succeed in obtaining good agreement within the given frequency uncertainties.

Our results therefore indicate that Hypothesis~2, which associates the radial modes with the frequencies at 64 and 71\cd, provides good agreement between observations and theory. This means that the dominant mode F10, which is located between the two radial overtones, is interpreted as a dipole mode.

\subsection{Detailed fit of the observed frequencies and examination of mode excitation}

\begin{figure}
  \centering
    \includegraphics[width=9cm,bb=10 15 287 187]{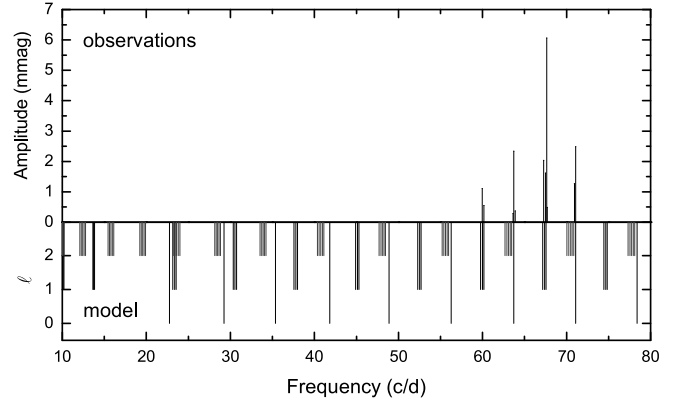}
  \caption{Comparison between observed frequencies and the theoretical frequency spectrum according to Model 2 given in Table \ref{tab:models}.}
  \label{fig:freqfit}
\end{figure}

As can be seen in Figure~\ref{fig:freqfit}, once the radial modes are fitted according to Hypothesis 2, the predicted dipole modes are in good agreement with the observed frequencies as well. We also find that the observed frequencies F8 (67.27\cd), F9 (67.50\cd), and F10 (67.64\cd) can be explained as a rotationally split triplet if an equatorial rotation rate of approximately 15 kms$^{-1}$ is assumed.

\begin{figure}
  \centering
    \includegraphics[width=9cm,bb=10 15 300 219]{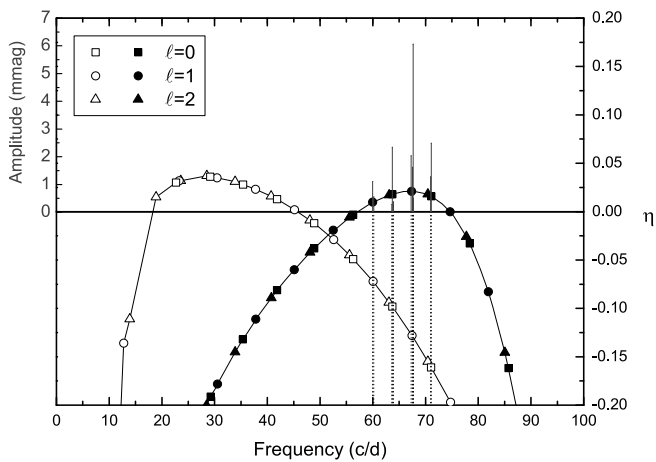}
  \caption{Instability parameter, $\eta$, for $\ell \leq 2$ modes in a model with solar composition (open symbols, Model 1 in Table~\ref{tab:models}) and in a model with modified metallicity and helium mass fractions (filled symbols, Model 2). $\eta>0$ for unstable modes. Vertical lines mark the position of observed frequencies.}
  \label{fig:eta}
\end{figure}

However, a nonadiabatic analysis shows that the observed pulsation modes are not predicted to be unstable by a model obtained with solar chemical composition (see Figure \ref{fig:eta}). Instead, instability is found at lower frequencies. To find a model that exhibits mode excitation in the observed frequency range, we conducted a mode instability survey based on a grid of fitted models, i.e. models that fit the observed frequencies according to Hypothesis 2. These models were obtained by varying different input parameters such as chemical composition and $\alpha_{\rm MLT}$. The mass of the model was no input parameter since it results from the fit of the theoretical frequencies to their observed counterparts. 

It turned out that the main effect on the mode instability of the fitted models is changes in He abundance and metallicity, while changing the efficiency of envelope convection through the mixing length parameter of convection, $\alpha_{\rm MLT}$, only has a marginal effect on mode instability in the given range of effective temperatures. 
For a subset of models obtained with different chemical composition, the frequency ranges of unstable modes and the frequency of the low-degree mode with strongest driving, $\eta_{\rm max}$, are presented in Table~\ref{tab:nadmodels}.
A change in chemical composition influences not only mode instability (which is determined by the conditions in the envelope) but also fundamental parameters, such as the effective temperature and luminosity (i.e., the position in the HR diagram). 
For solar chemical composition, the model is located at the ZAMS, while decreasing metallicity or increasing He abundance causes the fitted model gradually to become slightly more evolved. This also explains why in some cases, i.e., in models with high Z and high X in Table~\ref{tab:nadmodels}, no corresponding fitted model could be found on the main sequence.

The best agreement between predicted mode instability and observations could be achieved by a moderate change in chemical composition. Decreasing the metallicity from Z\,=\,0.0134 to 0.010 moves the maximum of the instability parameter, $\eta$, to higher frequencies but, at the same time, lowers the maximum in $\eta$ slightly. A slight increase in the He mass fraction, Y, to 0.29 increases the amount of driving and provides the best agreement between observations and theoretical predictions (see Figure \ref{fig:eta}).

\begin{figure}
  \centering
    \includegraphics[width=9cm,bb=10 10 323 240]{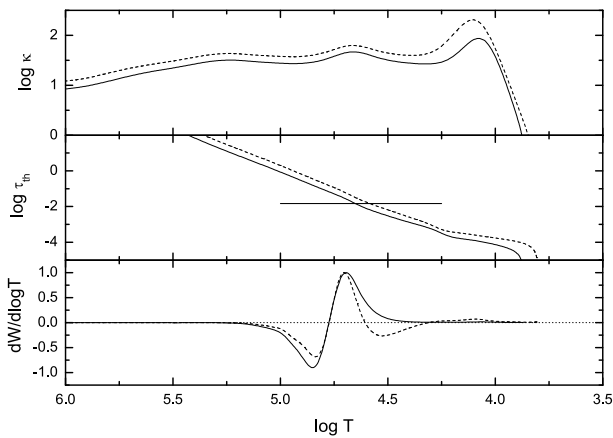}
  \caption{Rosseland mean opacity, $\kappa$, thermal time scale, $\tau_{\rm th}$ (in days) and differential work integral, d$W/$d$\log T$ (in arbitrary units), 
for the dipole mode at 67.6\cd. Model 1, in which this mode is damped, is shown with a dashed line, and Model 2, in which the mode is driven, by a solid line. The horizontal line in the middle panel depicts the period of the mode.}
  \label{fig:dWOpac}
\end{figure}

The differences in opacities and the differential work integral between the model obtained with solar chemical composition (Model 1 in Table~\ref{tab:models}) and the optimum model with X=0.70, Z=0.010 (Model 2) are given in more detail in Figure~\ref{fig:dWOpac}. Qualitatively, the behavior of the opacity is similar in both models. The main driving occurs in the inner part of the He~II ionization zone at $\log T$ about 4.6-4.75, where opacity is \emph{increasing} outwards. The outer part of the He~II ionization zone, where opacity is \emph{decreasing} outwards, potentially contributes to the damping of oscillations. This effect (i.e., negative values of differential work integral at $\log T$ around 4.5) can actually be seen in Model 1. In this model the cumulative work integral is negative which means that this oscillation mode is damped. For Model 2, which is hotter, the outer part of the He~II ionization zone at $\log T\approx4.5$ is located closer to the surface, where the thermal time scale is significantly smaller than the period of the dominant mode. Therefore the potential damping region tends to be in thermal equilibrium (neutral stability). The cumulative work integral is positive; i.e., the oscillation mode at 67.6\cd is excited.

\begin{table*}
 \caption{Results of a nonadiabatic study to match predicted mode instability with the observed frequencies. }
 \label{tab:nadmodels}
 \centering
 \begin{tabular}{llllllll}
  \hline
   \noalign{\smallskip}
     Opacity & X & Z & Mass ${[\mathrm{M}_\odot]}$ & log T$_{\rm eff}$ & log L/L$_\odot$ & Frequency range [\cd] & Frequency [\cd] \\
     &  &  &  &  &  &  with $\eta>0$  & of $\eta_{\rm max}$  \\
     \noalign{\smallskip}
     \hline
     \noalign{\smallskip}
     OPAL  & 0.76 & 0.0134 & -- & -- & -- & -- & -- \\
     OPAL  & 0.74 & 0.0134 & 1.576 & 3.874 & 0.757 & 16.8 - 55.3 & 23.3 \\
     OPAL  & 0.72 & 0.0134 & 1.604 & 3.894 & 0.844 & 24.6 - 54.4 & 37.8 \\
     OPAL  & 0.70 & 0.0134 & 1.623 & 3.911 & 0.915 & 38.2 - 61.0 & 52.5 \\
     OPAL  & 0.68 & 0.0134 & 1.615 & 3.921 & 0.951 & 42.5 - 68.0 & 56.3 \\
     OPAL  & 0.65 & 0.0134 & 1.599 & 3.928 & 0.979 & 45.3 - 72.9 & 60.1 \\
     \noalign{\smallskip}
     OPAL  & 0.74 & 0.0160 & -- & -- & -- & -- & --\\
     OPAL  & 0.74 & 0.0140 & 1.683 & 3.897 & 0.868 & 30.4 - 54.8 & 41.8 \\
     OPAL  & 0.74 & 0.0120 & 1.646 & 3.902 & 0.880 & 34.3 - 55.7 & 45.2 \\
     OPAL  & 0.74 & 0.0100 & 1.698 & 3.927 & 0.991 & -- & 60.0 \\
     OPAL  & 0.74 & 0.0008 & 1.718 & 3.945 & 1.069 & -- & 67.4 \\
     \noalign{\smallskip}
     OPAL  & 0.70 & 0.0160 & -- & -- & -- & -- & --\\
     OPAL  & 0.70 & 0.0140 & 1.604 & 3.903 & 0.878 & 30.5 - 57.3 & 45.2 \\
     OPAL  & 0.70 & 0.0120 & 1.647 & 3.925 & 0.974 & 47.6 - 68.3 & 60.1 \\
     OPAL  & 0.70 & 0.0100 & 1.662 & 3.941 & 1.039 & 56.8 - 74.4 & 67.4 \\
     OPAL  & 0.70 & 0.0008 & 1.670 & 3.956 & 1.102 & 69.3 - 75.4 & 71.2 \\
     \noalign{\smallskip}
     OP  & 0.74 & 0.0134 & 1.631 & 3.889 & 0.827 & 21.7 - 54.8 & 35.4 \\
     OP  & 0.74 & 0.010  & 1.741 & 3.936 & 1.034 & --          & 63.8 \\
     OP & 0.70 & 0.010  & 1.695 & 3.947 & 1.071 & 60.8 - 75.1 & 67.4 \\
     OP  & 0.70 & 0.011  & 1.694 & 3.941 & 1.045 & 56.3 - 73.4 & 67.4 \\
     \noalign{\smallskip}
     \hline
     \noalign{\smallskip}
      & & & & \multicolumn{2}{r}{observed} & 59.9-71.1 & 67.6 \\
   \hline
 \end{tabular}
 \tablefoot{
All models fit the four frequency groups following hypothesis 2 and were obtained with the A09 element mixture adopting Vrot=15.0 km/s. V$_{\rm rot}$ = 15.0 ${\mathrm{km\,s^{-1}}}$.}
\end{table*}

The detailed parameters of the best model (Model 2) are given in Table~\ref{tab:models} and the comparison between the observed and the predicted frequency spectrum is shown in Figure \ref{fig:freqfit}.
In the dipole groups of the predicted frequency spectrum, the thin substructure can be explained by rotational splitting. The substructure around the radial modes can be partly explained by rotationally split quadrupole modes, whose frequencies are close to those of radial modes. However, a detailed modeling of the observed substructures has not been performed, since this would require a better frequency resolution.

\begin{table}
  \caption{Fundamental parameters of two representative models.}
  \label{tab:models}
  \centering
  \begin{tabular}{l|ll}
   \hline
    \noalign{\smallskip}
      & Model 1 & Model 2\\
     X & 0.74   & 0.70 \\
     Z & 0.0134 & 0.010 \\
     Mass ${[\mathrm{M}_\odot]}$ & 1.576 & 1.662 \\
     log T$_{\rm eff}$ & 3.874 & 3.941 \\
     log L/L$_\odot$  & 0.757  & 1.04 \\
     R/R$_\odot$      & 1.428  & 1.456 \\
     log g           & 4.325 & 4.332 \\
     V$_{\rm rot}$ ${[\mathrm{km\,s^{-1}}]}$ & 15.0 & 15.0\\
    \noalign{\smallskip}
    \hline
  \end{tabular}
\end{table}

We repeated our nonadiabatic analysis using the older element mixture by Grevesse \& Noels (GN93) and had to adopt similar changes to the composition to find an optimum fit. 
Consequently, these changes in chemical composition and element mixture do not affect the frequencies of the modes in the fitted models much, but are important in modeling the outer envelope, where the mode instability is determined.

\section{Summary}

MOST high-precision time-series photometry was obtained for HD~144277 in 2009 and 2010, and its \dsct nature was discovered by MOST. The pulsation frequencies of HD~144277 lie between about 59.9 and 71.1\cd with amplitudes at the millimagnitude level. Four equidistant groups of frequencies were identified with spacings of about 3.6\cd.

The observed high frequencies in this hot A star indicate that we are in the asymptotic region of the frequency spectrum. The spacings of 3.6\cd\,\, therefore correspond to the spacing between subsequent radial and dipole modes. With a resulting radial spacing of 7.2\cd\,\, between radial modes, HD~144277 represents a \dsct star close to the ZAMS.

Despite the lack of reliable fundamental parameters, e.g. no measured $v\sin i$ value, we attempted an asteroseismic interpretation of the observed pulsation spectrum. As discussed in section 4.1, rapid rotation cannot be ruled out for this star and it may be worthwhile examining it. In this paper, however, we restrict ourselves to interpreting the spectrum in the framework of slow rotation for which our codes are valid. A high-resolution spectrum for HD~144277 will be obtained in the near future for more observational and theoretical analyses.


We conducted an asteroseismic analysis of radial period ratios to identify the spherical degree of the dominant peaks in the spectrum. We found evidence that two radial modes, the sixth and seventh overtones could be assigned to frequency group B at 63\cd and group D at 70\cd.
The mode with the highest amplitude, F10 at 67.6\cd, would then be identified as a dipole mode. Together with the nearby modes F9 (67.50\cd) and F11 (70.96\cd), it forms a substructure that we interpret as rotational splitting. The derived equatorial rotational velocity amounts to 15 kms$^{\rm -1}$.

In the case of models obtained with standard chemical composition, mode instability is incorrectly reproduced; instead, a reduction in metallicity (from Z\,=\,0.0134 to 0.010) and an increase in helium abundance (from Y\,=\,0.2466 to 0.29) has to be adopted to fit the observed instability range.  In fact, the reduction of metallicity implies that the observed frequencies are fitted by models at higher effective temperature, where higher order modes are driven. If the assumption about slow rotation holds true, then we are able to test the physical conditions of the envelope within the frame of the existing atomic physics data in HD~144277.

\begin{acknowledgements}
We are grateful to Gerald Handler, who obtained the Stroemgren photometry for HD~144277 for us at the SAAO 50cm telescope.

KZ is recipient of an APART fellowship of the Austrian Academy of Sciences at the Institute of Astronomy of the University of Vienna. PL, AAP, and TZ acknowledge partial financial support from the Polish MNiSW grant No. N~N203 379 636. MB and WWW acknowledge support by the Austrian {\it Fonds zur F\"orderung der wissenschaftlichen Forschung} (MB: project P21830-N16; WWW: project P22691-N16).
\end{acknowledgements}


\begin{thebibliography}{}
  \bibitem[2009]{asplund09} Asplund, M., Grevesse, N., Sauval, A. J., Scott, P. 2009, ARA\&A, 47, 481
\bibitem[1993]{bre93} Breger, M., Stich, J., Garrido, R., et al. 1993, \aap, 271, 482
\bibitem[2008]{breger08} Breger, M., Lenz, P. \& Pamyatnykh, A. A. 2008, CoAst, 157, 56
  \bibitem[2009]{breger09} Breger, M., Lenz, P. \& Pamyatnykh, A. A. 2009, MNRAS, 396, 291
  \bibitem[2011]{breger11} Breger, M., Balona, L., Lenz, P., et al. 2011, \mnras, in press
\bibitem[2011]{burke2011} Burke, K. D., Reese, D. R., Thompson, M. J. 2011, MNRAS, 414, 1119
  \bibitem[2005]{ferg05} Ferguson, J. W., Alexander, D. R., Allard, F., et al. 2005, ApJ, 623, 585
  \bibitem[1993]{gn93} Grevesse, N., \& Noels, A. 1993, in Prantzos N., Vangioni-Flam E. and Casse M. eds., Origin and Evolution of the Elements, CUP, p.15
\bibitem[2008]{har08} Hareter, M., Reegen, P., Kuschnig, P., et al. 2008, CoAst, 156, 48
\bibitem[2004]{espinosa2004} Espinosa, F., Pérez Hernández, F., Roca Cortés, T. 2004, in Proceedings of the SOHO 14 / GONG 2004 Workshop (ESA SP-559). "Helio- and Asteroseismology: Towards a Golden Future", Vol. 559, p.424
  \bibitem[1996]{iglesias1996} Iglesias, C. A., \& Rogers, F. J. 1996, A\&A, 464, 943
\bibitem[2005]{len05} Lenz, P. \& Breger, M. 2005, CoAst, 146, 5
  \bibitem[2010]{lep2010} Lenz, P., Pamyatnykh, A. A., Zdravkov, T., Breger, M. 2010, A\&A, 509, A90
\bibitem[2009]{lig09} Lignieres, F. \& Georgeot, B. 2009, \aap, 500, 1173
\bibitem[2010]{lig10} Lignieres, F., Georgeot, B. \& Ballot, J. 2010, AN, 331, 1053
\bibitem[2008]{kal08} Kallinger, T., Reegen, P. \& Weiss, W. W. 2008, \aap, 481, 571
\bibitem[2009]{kha09} Kharchenko, N.V. \& Roeser, S. 2009, Kinematics and Physics of Celestial Bodies, 17, 409
\bibitem[1997]{kus97} Kuschnig, R., Weiss, W. W., Bahr, R., Bely, P., Jenkner, H., 1997, A\&A, 328, 544
\bibitem[2000]{pam00} Pamyatnykh, A. A. 2000, in Delta Scuti and Related Stars, Reference Handbook and Proceedings of the 6th Vienna Workshop in Astrophysics, ASP Conf. Series, Vol. 210, Michel Breger and Michael Montgomery, eds., San Francisco: ASP, p215
  \bibitem[1972]{pet1972} Petersen, J. O. \& J{\o}rgensen, H. E. 1972, A\&A, 17, 367
\bibitem[2006]{ree06} Reegen, P., Kallinger, T., Frast, D., et al. 2006, \mnras, 367, 1417
\bibitem[2007]{ree07} Reegen, P. 2007, \aap, 467, 1353
\bibitem[2009]{rees09} Reese, D. R., MacGregor, K. B., Jackson, S., Skumanich, A., Metcalfe, T. S 2009, \aap, 506, 189
\bibitem[2005]{seaton05} Seaton, M. J. 2005, MNRAS, 362, L1
\bibitem[2003]{wal03} Walker, G., Matthews, J. M., Kuschnig, R., et al. 2003, \pasp, 115, 1023
\bibitem[2003]{wri03} Wright, C. O., Egan, M. P., Kraemer, K. E., Price, S. D. 2003, \aj, 125, 359
\bibitem[2009]{zwi09} Zwintz, K., Hareter, M., Kuschnig, R., 2009, et al. \aap, 502, 239
\end{thebibliography}
\end{document}